\documentclass[]{aa501}
\usepackage{psfig}
\usepackage{graphicx}

\def\Ni{\noindent}

\def\etal{{et\,al.}}

\def\ergcm{\hbox{erg cm$^{-2}$ s$^{-1}$ }}
\def\erga{\hbox{erg cm$^{-2}$ s$^{-1}$ \AA$^{-1}$ }}

\def\grad{$^\circ$}
\def\it{\sl}
\def\degs{\ifmmode ^{\circ}\else$^{\circ}$\fi}
\def\amin{\ifmmode ^{\prime}\else$^{\prime}$\fi}
\def\asec{\ifmmode ^{\prime\prime}\else$^{\prime\prime}$\fi}
\def\fss{\hbox{$.\!\!^{\rm s}$}}        %
\def\h{$^{\rm h}$}
\def\m{$^{\rm m}$}

\def\rxj{RX\,J1846.9+5538}
\def\117{RX\,J1846}

\begin{document}

 \title{A new two-pole accretion polar: \rxj\thanks{Partly based on observations
         collected at the German-Spanish
          Astronomical Centre, Calar Alto, operated by the MPI f\"{u}r
          Astronomie, Heidelberg, jointly with the Spanish National
          Commission for Astronomy.}}

   \author{R. Schwarz\inst{1} \and
           J. Greiner \inst{1,2}
          \and G.H. Tovmassian\inst{3}, %
          S.V. Zharikov\inst{3}
          \and W. Wenzel\inst{4}
         }

  \authorrunning{R.~Schwarz et al.}
   \offprints{R.Schwarz, rschwarz@aip.de}
 
  \institute{Astrophysical Institute
          Potsdam, An der Sternwarte 16, 14482 Potsdam, Germany
        \and Max-Planck-Institut f\"{u}r Extraterrestrische Physik,
           Giessenbachstra\ss e, D--85740 Garching, Germany
       \and Instituto de Astronom\'{\i}a, UNAM, Apdo.Post. 877, 22860
            Ensenada, B.C., M\'{e}xico       
       \and Sonneberg Observatory, D--96515 Sonneberg, Germany
      } 

   \date{Received 9 April 2002/ Accepted 29 May 2002}
   \abstract{
We report the discovery of a  %
new, bright ($V\sim 17^{\rm m}$) AM Her system as the optical counterpart of the
soft ROSAT All-Sky-Survey source \rxj\ (= 1RXS J184659.4+553834). 
Optical photometric and spectroscopic follow-up observations reveal a single 
period of 128.7 min, consistent with a high degree of spin-orbit 
synchronization, and a
low polar field strength ($B<20$~MG) of the primary accretion region. 
The system was observed in optical intermediate and high states that 
differ by about 1 mag. 
These  brightness variations were accompanied by a 
correlated change of the optical light curve, which we interpret as a switch 
between one- and two-pole accretion.
This 
explanation
is also supported by the X-ray light curves,
which at two different 
epochs display emission from two equally bright accretion regions
separated by $\sim 160\degs$. Both spots possess distinct spectral 
X-ray properties
as seen from the X-ray hardness ratio, where the secondary accretion region
appears significantly softer, thus probably indicating a higher field strength 
compared to the primary region.
In all ROSAT pointings  a deep dip is present during the primary flux maxima, 
very likely caused by absorption in one of the accretion streams. 
}
\maketitle
\section{Introduction}

AM Her type variables  are a subgroup of cataclysmic variables
in which the magnetic field of the white dwarf controls the geometry 
of the material flow between the main-sequence donor and the white dwarf
primary (see e.g. Warner 1995 for a detailed review). 
The inflow of matter along the magnetic field lines (of one or
occasionally also two magnetic poles) is decelerated above the white dwarf
surface producing a shock front. This region is thought to emit hard X-rays
(usually modelled in terms of thermal bremsstrahlung of 10--20 keV)
and polarized cyclotron radiation (hence these systems are also named polars) 
in the IR to UV range. 
In addition, a strong soft component has been frequently observed from polars
that is thought to arise from the heated accretion pole
(usually modelled in terms of a blackbody of 20--50 eV).

  \begin{figure}[t]
       \fbox{\includegraphics[clip,width=0.95\columnwidth]{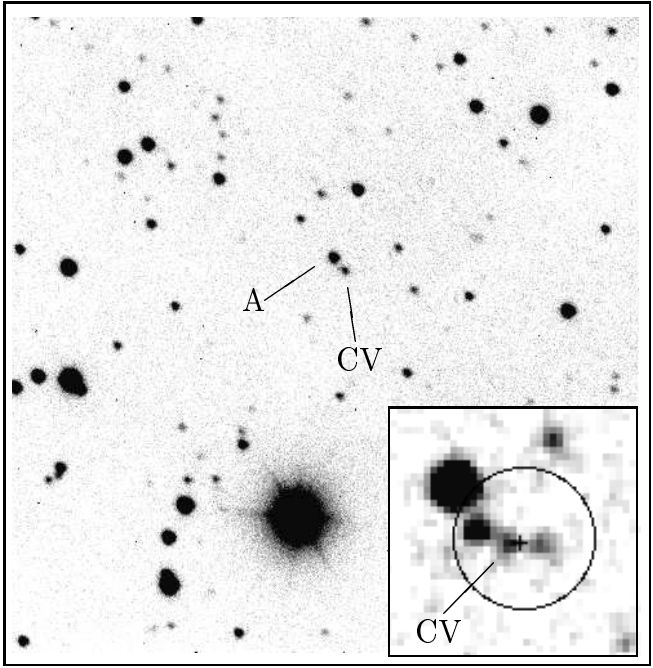}}\par
      \caption[fc]{White-light CCD image of \rxj\ obtained with the
        AIP 70 cm telescope. North is top and East to the left.
         The size of the field is approximately 8\amin$\times$8\amin.
         The inset at the lower right corner shows a 30\asec$\times$30\asec\ 
         blow up taken from the digitized Palomar Observatory Sky Survey 
         (POSS) 2 together with the 2$\sigma$ X-ray error circle derived 
         from the HRI pointing. The bright star in the upper left of the
         blow up is star A.
         The cataclysmic variable \rxj\ was certainly in an optically low state
         during the POSS exposure. Its optical position was determined to
         $\alpha_{2000}$ = 18\h46\m58\fss9 and 
         $\delta_{2000}$ = 55\grad 38\amin 29\asec\ ($\pm$1\asec).
          }
      \label{fc}
\end{figure}

It is this soft X-ray component which has led to the discovery of a few
dozen new polars by ROSAT observations over the last decade, 
most notably the ROSAT all-sky survey (Beuermann \& Burwitz 1995). 
The source described here has been
discovered as a result of a systematic survey for supersoft X-ray sources from 
the all-sky survey data (Greiner 1996 for details of this survey)
which revealed a large number of CVs and single white dwarfs. 
Other confirmed polars identified from this sample include 
V844 Her = RX\,J1802.1+1804 (Greiner, Remillard and Motch 1995, 1998),
RS Cae = RX\,J0453.4--4213 (Burwitz \etal\ 1996) and 
V1007 Her = RX\,J1724.0+4114 (Greiner, Schwarz and Wenzel 1998).

In this paper we present photometric, spectroscopic
and X-ray observations (summarized in Table~\ref{log} and \ref{xlog})
that led to the discovery of a new polar, \rxj\ 
(henceforth referred to as \117).

\section{ROSAT X-ray observations}
 \begin{table} 
\caption{Log of optical observations}
\begin{tabular}{lllrr}
      \hline
      \noalign{\smallskip}
 Telescope  & ~~~Date  &  Range & $T_{\rm Dur}$ & $T_{\rm Int}$ \\
 && (\AA/ Filter) & (hrs) & (sec) \\
      \noalign{\smallskip}
      \hline
      \noalign{\smallskip}
 \multicolumn{5}{c}{spectroscopy} \\
 CA 3.5 m & 1992 Oct 1   & 3900--7200 & 1.0 & 3600 \\
 SAO 6.0 m & 1999 Jun 19 & 3400--7900 & 3.6 & 900 \\
   \noalign{\smallskip}
 \multicolumn{5}{c}{photometry} \\
 SO 0.6 m & 1994 Sep 21 & $R$ & 4.1 & 600 \\
 SO 0.6 m & 1994 Sep 22 & $R$ & 3.7 & 600 \\
 SO 0.6 m & 1994 Sep 23 & $R$ & 2.2 & 600 \\
 SO 0.6 m & 1994 Sep 25 & $R$ & 0.5 & 600 \\
 SO 0.6 m & 1994 Sep 27 & $W\!L$ & 0.9 & 300 \\
 SO 0.6 m & 1994 Sep 29 & $W\!L$ & 4.8 & 180 \\
 SO 0.6 m & 1994 Oct 12 & $R$ & 1.3 & 600 \\
 SO 0.6 m & 1994 Oct 13 & $W\!L$ & 0.4 & 600 \\
 SO 0.6 m & 1994 Oct 14 & $W\!L$ & 0.3 & 600 \\
 OAN 1.5 m & 1995 May 2  & $W\!L$ & 3.2 & 120 \\
 OAN 1.5 m & 1995 May 4  & $W\!L$ & 2.6 & 90 \\
 AIP 0.7 m & 1995 Oct 24 & $W\!L$ & 2.3 & 120 \\
 AIP 0.7 m & 1997 Oct 28  & $W\!L$  & 1.3 & 60 \\
 AIP 0.7 m & 1998 Jan 25  & $W\!L$  & 4.7 & 30 \\
 AIP 0.7 m & 1998 Mar 24  & $W\!L$  & 3.7 & 60 \\
 AIP 0.7 m & 1998 Aug 5   & $W\!L$  & 3.3 & 60 \\  
 AIP 0.7 m & 1998 Sep 24  & $W\!L$  & 2.3 & 60\\  
 \noalign{\smallskip}
 \hline
 \noalign{\smallskip}
 \end{tabular}

\noindent{\small The abbreviations have the following meaning:
  CA = Calar Alto,
  SAO = Special Astrophysical Observatory, Zelenchukskaja,
  SO = Sonneberg Observatory,
  OAN = Observatorio Astron\'omico Nacional de San Pedro M\'artir, 
  AIP = Astrophysical Institute Potsdam,
  $W\!L$ = white light
  }
   \label{log}
 \end{table}

\begin{table}
\caption{Log of the X-ray observations of \rxj}
\begin{tabular}{ccccc}
\noalign{\smallskip} \hline \noalign{\smallskip}
      Date$^{(1)}$ & $T_\mathrm{Exp}$ & N$_{\rm cts}$ & Mean rate & $H\!R1$ \\
                 &          &   & (cts/s) & \\
\noalign{\smallskip} \hline \noalign{\smallskip}
    1990 Oct 12--19~P  & ~1610 & ~\,136 & 0.12   & $-0.85$ \\
    1992 Jun 15--18~P  & 10540 & 2478   & 0.28   & $-0.74$ \\
    1993 Sep 25/26~P   & 27510 & 1895   & 0.10   & $-0.60$ \\
    1995 Apr 08--13~H  & 17950 & ~\,280 & ~~~0.02$^{(2)}$ &  --   \\
 \noalign{\smallskip}
 \hline
 \noalign{\smallskip}
      \end{tabular}\\
   \noindent{\Ni\small $^{(1)}$ The letters after the date denote the
   ROSAT detector: H = HRI, P = PSPC\\ 
                       $^{(2)}$ Note  the lower sensitivity of the
                                HRI at soft energies by a factor of 7.8.
            }
    \label{xlog}
   \end{table}

\subsection{All-Sky Survey and optical identification}\label{s:rx1846_rass}
\begin{figure}[th]
      \includegraphics[width=0.98\columnwidth,clip]{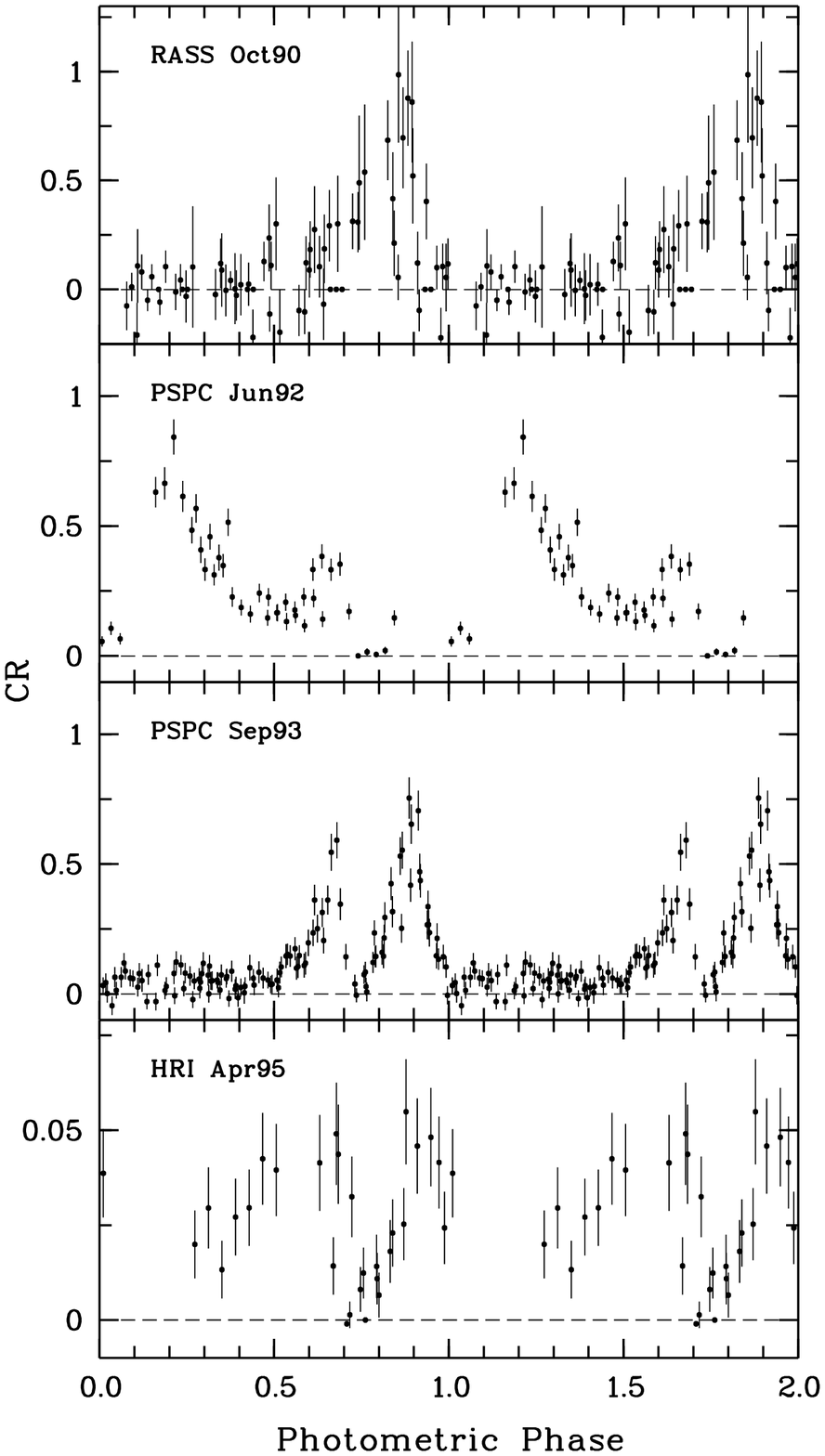}
      \caption[fc]{ROSAT X-ray light curves of \rxj .  }
      \label{xlc}
\end{figure}

\117\ was scanned during the ROSAT All-Sky-Survey (RASS) over a period of 8 days
in October 1990 for a rather long (due to its high ecliptic latitude) 
total observing time of 1610 sec. Its mean count rate in the ROSAT 
position-sensitive proportional counter (PSPC) was 0.2 cts/s,
and the hardness ratio $H\!R1=-0.85\pm0.06$, where $H\!R1$ is defined as 
(H--S)/(H+S), with H (S) being the counts above (below) 0.4 keV over
the full PSPC range of 0.1--2.4 keV.

Within the X-ray positional error circle (see  Fig.~\ref{fc})
there are two bright optical sources, 
USNO~A2.0 1425.09302300  (named Star A in the finding chart) 
and 12\arcsec\ SW of that, the cataclysmic binary identified spectroscopically
(see Sect.~\ref{s:rx1846_spec}).
Its optical position is measured as: R.A. (2000.0) = 18\h46\m58\fss9,
Decl. (2000.0) = 55\grad 38\amin 29\asec\ ($\pm$1\asec).
Inspection of the digitized Palomar Observatory Sky Survey 2 plates taken 
in October 1991 (see blow-up in the lower right 
corner of Fig.~\ref{fc}) revealed two additional fainter sources 
5\arcsec\ east and west of \117. At that time the CV had dropped 
into a deep low state of $\sim 20^{\mathrm m}$.

\subsection{X-ray light curves}
For the timing analysis the RASS photons were extracted within
a radius of 4\amin\ around the X-ray position of \117 .
The background was chosen at the same ecliptic longitude at
$\sim$1\degs\ distance, corresponding to the photons detected
typically 15 sec before or after the arrival of the source photons. 
Standard corrections were applied using the dedicated EXSAS software package
(Zimmermann et al. 1994) including barycenter correction.

Further dedicated follow-up pointed ROSAT observations were 
performed on June 15--18,
1992 with the PSPC and on April 8--13, 1995 with the high-resolution imager
(HRI). In addition, \117\ was also covered in a serendipitous pointing on
September 25/26, 1993 at about 37\amin\ off-axis angle. 
Using the EXSAS package, source photons were extracted with a radius of
1\farcm5, 7\farcm5 and 0\farcm8 for the on-axis PSPC pointing, the off-axis
PSPC pointing and the HRI pointing, respectively. The background was chosen 
from concentric circles around the source region with radii of 3\amin,
12\farcm5 and 2\amin, respectively. Other nearby sources were cut out, and the
background area normalized to the source extraction area before the 
background subtraction.   
Table~\ref{xlog} summarises these measurements.

The RASS light curve 
folded over the photometric ephemeris as derived in
Sect.~\ref{s:orb} is shown in  Fig.~\ref{xlc} (upper panel).
It displays   a 100\% modulation with a peak count rate of nearly 1
cts/s and a pronounced faint-phase where the X-ray flux is practically
zero (formal count rate of $-0.006\pm 0.10$ cts/s). 
This `on-off' profile is typical for polars where the emission 
region disappears behind the limb of the white dwarf for part of the orbit. 
The short duration of the bright phase, which lasts only 0.3 of the
orbit, requires the main accretion region to be in the 
lower hemisphere of the white dwarf. 
A similar morphology was also seen during the longest pointed observation
in September 1993, but now with a bright phase extended to $\Delta\phi 
\sim 0.5$. 
The phasing of that X-ray bright phase strongly suggests that it is due
to the same primary accretion region seen also in the optical light curves
during the intermediate state in 1997/98.
The PSPC observation in September 1993 reveals a non-zero count rate 
($0.048\pm 0.002$) also during the faint phase. 

This simple 'bright-faint' behaviour was fundamentally altered
in the June 1992  PSPC pointing. 
Most strikingly, we observe the X-ray flux  peaking at $\phi = 0.23$, 
the phase of the former faint interval. Although the rise to maximum is not 
covered and the decline probably overlaps with emission from the still active 
primary region, the shape of this hump is suggestive of emission from a 
second self-eclipsing accretion region. This is even more convincingly 
confirmed by the spectral softening during the flux peak 
(see Sect.~\ref{s:rx1846_xspec}).
This secondary accretion spot is equally bright compared to the primary one,
reaching a count rate of 0.8 cts/sec.

Further HRI observations were carried out in April 1995, which together 
with the photometry obtained three weeks later provide the closest 
X-ray/optical observations in time.
The lack of complete orbital coverage and spectral resolution render these 
observations not completely conclusive. 
At that time the system was also X-ray bright at phase 0.3, which 
might be interpreted as two-pole accretion, probably seen also later in the
optical. On the other hand, the flux peak at phase 0.25 as seen in June 1992 
is missing, and the light curve can also be explained as a result of a 
prolonged primary bright phase. 

All pointed observations show, contrary to the RASS observation, 
a marked decrease of the X-ray flux in the bright phase at around $\phi
\sim 0.7$ .  During the 1993 PSPC and the 1995 HRI observation the dip 
had a short ingress, and was followed, after a 
short interval ($\Delta\phi <0.01$) of totality, by a rather protracted egress.
The duration of the dip at these occasions, 
from start to complete recovery of the pre-dip level, 
lasted  0.19 phase units. 
The shape of the dip was considerably altered in 1992, 
when it was more flat-bottomed and the X-ray flux was zero for about 
$\Delta\phi \simeq 0.07$.  
Dip egress was insufficiently covered in this pointing.  
The occurrence of such dips in polars is in general understood as a 
consequence of photoelectric absorption by matter of the accretion stream. 
The X-ray spectral hardening during dip phase and the subsequent softening
during egress (see Sect.~\ref{s:rx1846_xspec}) are in line with the expected 
energy dependence of cold absorption. 
Possible alternative explanations, like an eclipse by the 
secondary or a highly structured accretion region, are not
consistent with the phase variability and the sharp ingress/egress 
of the dip observed in \117 . In context with the photoelectric absorption, 
the modification of the 
dip profile in the 1992 PSPC observation  would imply a denser and 
more collimated stream, supposedly connected with the increased activity 
of the secondary pole at that time. 

\subsection{X-ray spectroscopy}\label{s:rx1846_xspec}
\begin{figure*}
  \mbox{
  \psfig{figure=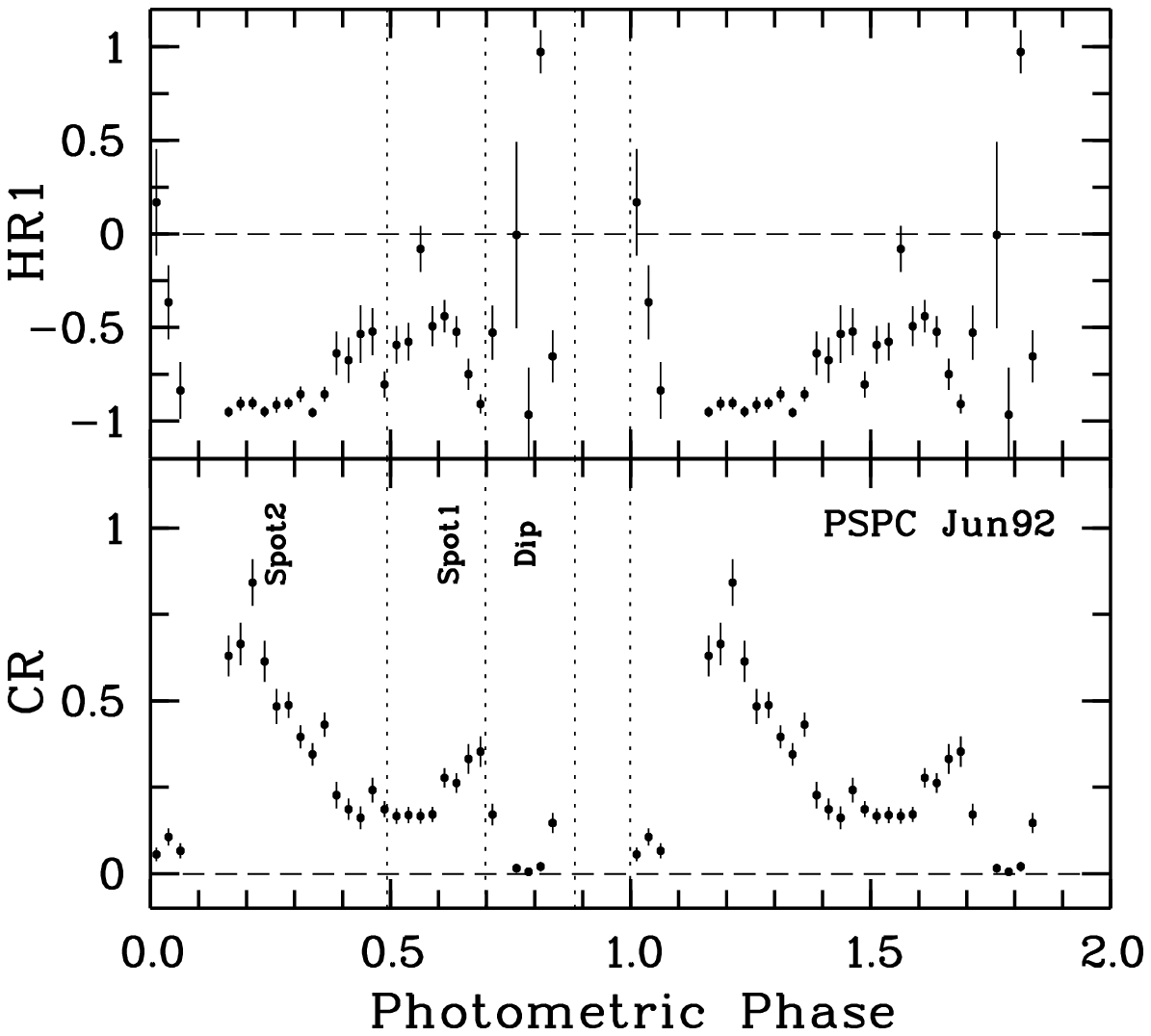,width=0.95\columnwidth,clip=}
  \hfill
  \psfig{figure=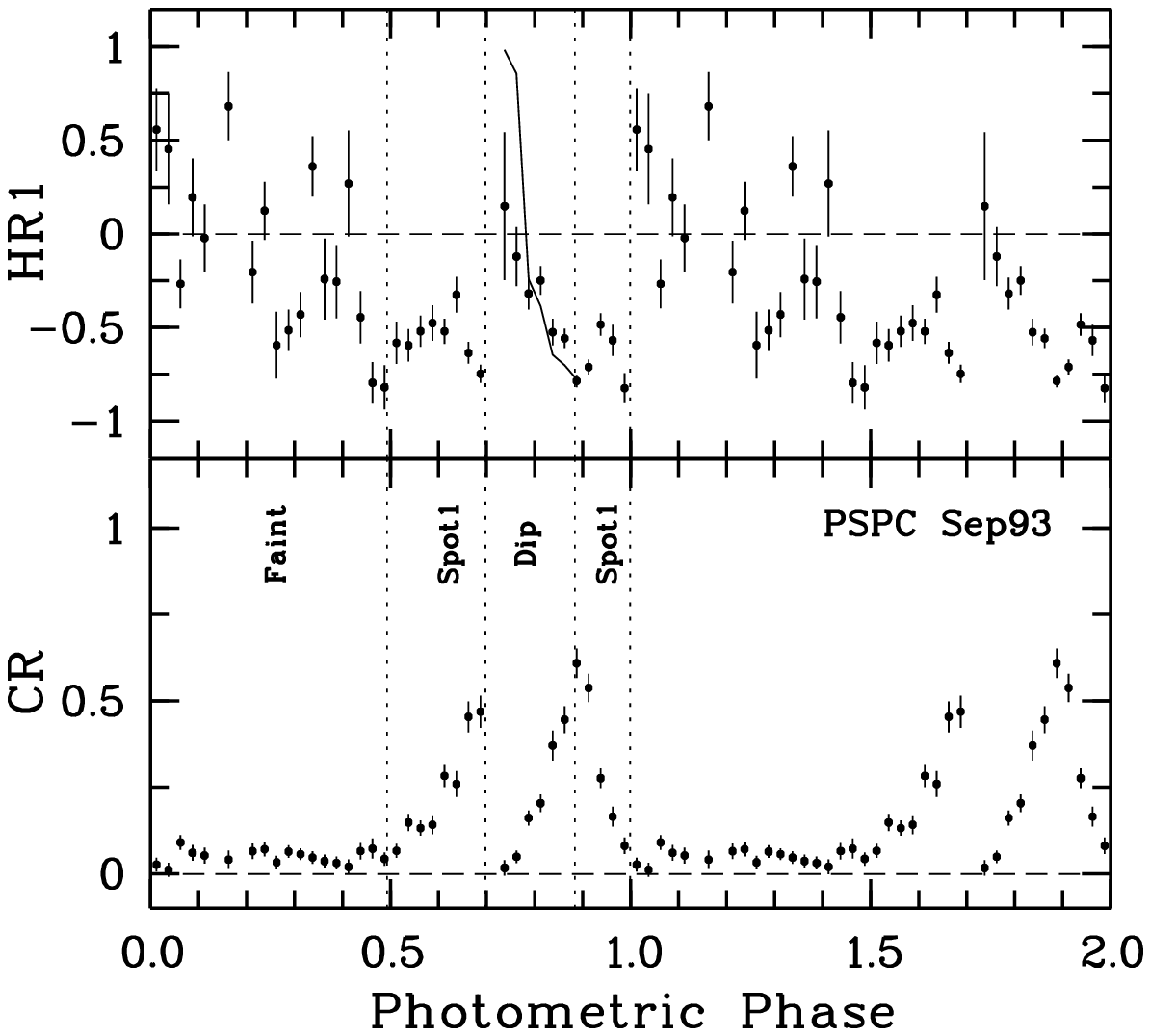,width=0.95\columnwidth,clip=}
  }
  \caption[hr1]{Hardness ratio variations (upper panel) of \117\ during 
  the pointed ROSAT observation in June 1992 (left) and September 1993 (right).
  In the lower panel the corresponding X-ray light curves are shown.
  Dotted vertical
  lines indicate the beginning and the end of the primary bright phase and 
  the absorption dip in the 1993 data. The solid line in the upper right 
  panel shows the predicted variation of $H\!R1$ in the absorption dip.}
  \label{f:rx1826_hr1}
\end{figure*}

   \begin{table}
     \caption{Fit results for phase-resolved X-ray spectra}
      \begin{tabular}{lcccc}
      \hline
      \noalign{\smallskip}
      Data & $N_{\rm H}$ 
                  & $F_{\rm bb}$ 
                  & $F_{\rm br}$ 
                  & $\chi^2_{red}$ \\
         &  (10$^{20}$ cm$^{-2}$)  & 
         \multicolumn{2}{c}{($10^{-12}$erg/cm$^{2}$/s)}
               &  \\
       \noalign{\smallskip}
      \hline
      \noalign{\smallskip}
       92 spot1 & 1.3$\pm$0.7 & 6.5  & 5.0 & 0.27 \\
       93 spot1 & 1.2$\pm$0.5 & 6.8  & 4.2 & 0.51 \\
       92 spot2 & 0.8$\pm$0.3 & 9.9  & 2.0 & 2.13 \\
       93 faint & 0.7$\pm$1.6 & 0.5  & 1.7 & 1.13 \\
      \noalign{\smallskip}
      \hline
      \noalign{\smallskip}
  \end{tabular}

\noindent{\small The blackbody and bremsstrahlung temperatures have been
fixed at $kT_{\rm bb}= 25$ eV and
$kT_{\rm br}= 20$ keV during the fit.  
                 The total galactic absorbing column is 
                 4.9$\times$10$^{20}$ cm$^{-2}$  (Dickey \& Lockman 1990).}
   \label{fitres}
   \end{table}

The two pointed PSPC observations have resulted in the detection of more
than 4400 photons, thus enabling spectral investigations along the orbital
period. As a first step we used the hardness ratio $H\!R1$ (as defined in Sect.~\ref{s:rx1846_rass})
which is shown as a function of the orbital period in Fig.~\ref{f:rx1826_hr1}.
It is highly variable, changing from $0.5$ to $-1$ depending on which accretion 
region was visible. For the primary accretion region that was covered during
both pointings the $H\!R1$ was consistently at an intermediate value 
of $-0.57$. 
Contrary to that, the secondary pole active only in June 1992 
($\phi = 0.1-0.35$) is supersoft with $H\!R1 = -0.92$. For 
the faint phase observed in the 1993 pointing the $H\!R1$ changes 
between $-0.5$ to $0.5$ with an average value of  $-0.11$, notably 
harder than the two main emission regions.  

The average X-ray spectrum of \117\ is well represented 
by a two component model consisting of a blackbody plus a thermal 
bremsstrahlung model having a low absorption ($N_{H}< 10^{21}$
cm$^{-2}$) and wide range of possible blackbody temperatures $15-40$ eV
(Fig. \ref{oxspec}). 
Since the temperatures of both components are not well constrained by the
ROSAT data and the photon statistics of the phase-resolved spectra are
low, we concentrate in the following on the determination of the X-ray fluxes 
assuming fixed values for the temperatures,  set to $kT_{\rm bb}= 25$ eV and
$kT_{\rm br}= 20$ keV. Spectra from the different emission regions
were extracted separately using the phase intervals $0-0.4$ (spot2 1992),
$0-0.492$ (faint phase 1993) and $0.492-0.697$, $0.492-0.697$ for the
primary accretion region.   
The resulting best-fit values of the absorption and the 
unabsorbed, bolometric blackbody and bremsstrahlungs fluxes 
are summarized in Table~\ref{fitres}. 
As already indicated by the count rates and hardness ratio, 
both accretion regions emit the same total X-ray flux of  
$F_{\rm X} \sim 1\times$10$^{-11}$ erg/cm$^2$/s, but have 
different flux ratios  
$F_{\rm thbr}$/$F_{\rm bbdy}$ of  0.6 and 0.02, for the primary and
secondary region, respectively. 

The luminosity observed during the faint phase in 1993 is a factor of 5
lower compared to primary accretion sites, and emission from that 
accretion area shows a marked soft X-ray deficiency. 
Such an effect has already been noted in the
ROSAT spectra of polars observed in intermediate and low states (Ramsay et al.
1995), and is likely caused by the dominance of cyclotron cooling 
in low accretion rate plasmas.

The dip phase is accompanied by a marked increase of the hardness ratio 
in both PSPC pointings.
In 1992, when the statistics were low, 
the mean $H\!R1$ was roughly zero. For the longer PSPC pointing where the dip 
was only total for a short interval, the $H\!R1$
also dropped to zero at the time of flux minimum, but then showed a correlated 
decrease as the X-ray count rate recovered to the pre-dip value.
We model the change of the $H\!R1$  under the assumption that 
it is caused by the absorption of neutral, cold matter.  For this purpose
we adopted a blackbody/thermal bremsstrahlung model with fixed 
$kT_{\rm bb}= 25$ eV and $kT_{\rm br}= 20$ keV for the intrinsic X-ray
source, and set their normalizations according to the count rate and
hardness ratio observed right after the dip egress. We then determined
what absorbing column is necessary to reduce the intrinsic X-ray flux
to the observed values, and computed the corresponding change of $H\!R1$.
As shown in Fig.~\ref{f:rx1826_hr1} (right) the cold absorption model 
reproduces the $H\!R1$ variation except for the lowest count rate bins,
where the observed $H\!R1$ = 0 is significantly softer than the predicted
value of 1. This difference might indicate a more complex absorbing medium 
such as partial covering or a warm absorber, or is simply due to a change of the 
intrinsic X-ray spectrum. 
The column densities within the accretion stream as derived from our 
modelling range from $1.1\times 10^{20}$~cm$^{-2}$ at the start and end
of the dip to $5\times 10^{21}$~cm$^{-2}$ in the most opaque parts.

\section{Optical photometry}

\117\ was monitored during 17 nights between  1994 and 1998 utilizing 
telescopes at Sonneberg Observatory,   
the Astrophysical Institute Potsdam at Babelsberg (both Germany)
and the Observatorio Astronomico Nacional, (Mexico) at San Pedro Martir.

These telescopes have apertures of 0.6~m, 0.7~m and 1.5~m, respectively, 
and were equipped with CCD detectors. Exposures varied from 30 sec
to 600 sec and were mainly taken without filters. Differential 
magnitudes have been computed with respect to star A (Fig.~\ref{fc}),
which has a $R$ magnitude of 15\fm6 as measured in the USNO catalog.
The images itself were processed using the profile-fitting scheme of the 
{\sc DoPhot} reduction package (Mateo \& Schechter 1989).
The combined log of all observations is given in Table~\ref{log}.

\subsection{Orbital variations}\label{s:orb}
\begin{figure*}[t]
   \psfig{figure=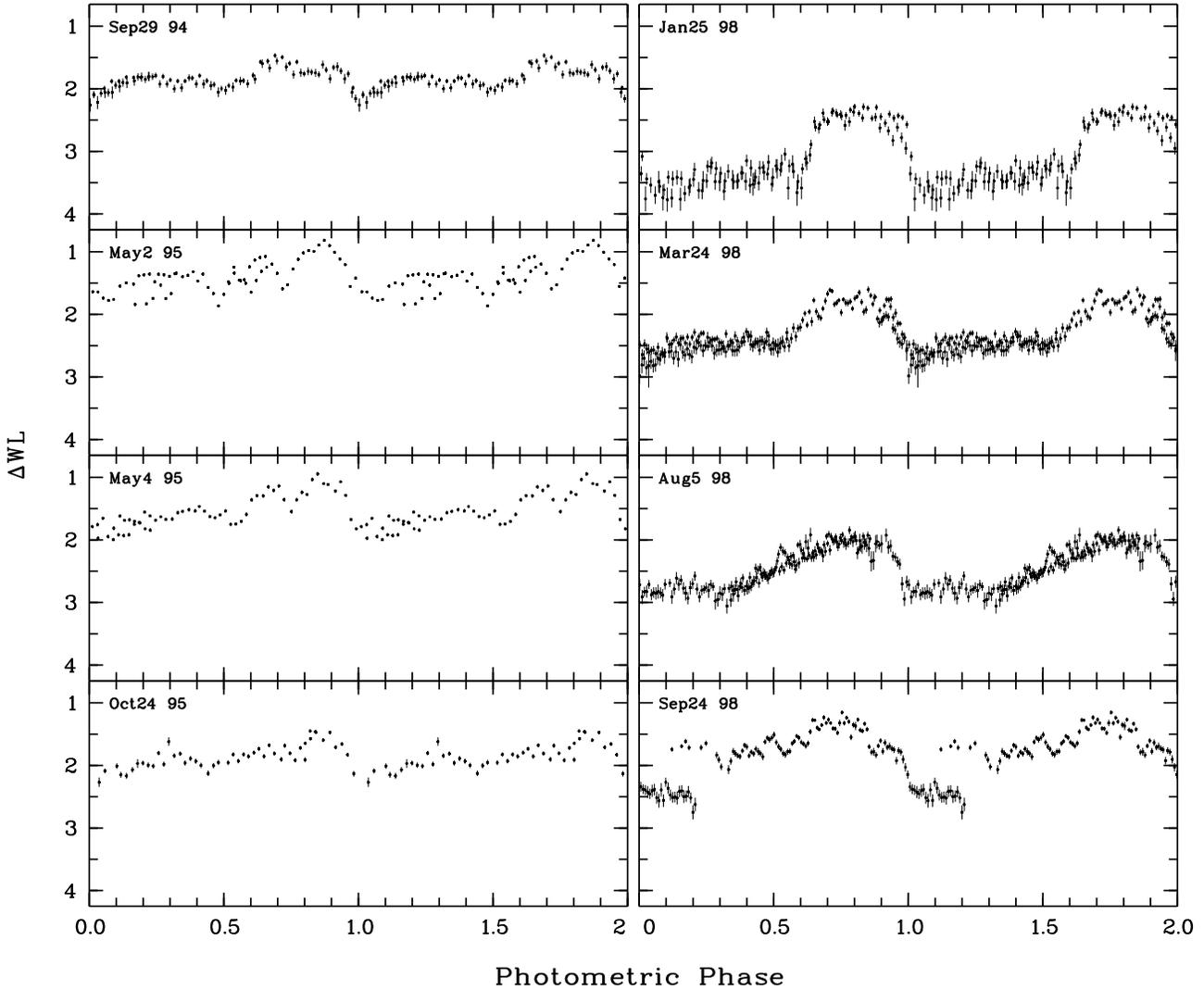,width=1.98\columnwidth}
   \caption[olc]{
   Optical light curves of \117\ plotted as a function of ephemeris
   Eq.~\ref{e:rx1846_eph}. Data are plotted twice for clarity.}
  \label{f:rx1846_lc}
\end{figure*}
\begin{figure}[th]
      \includegraphics[width=0.98\columnwidth,clip]{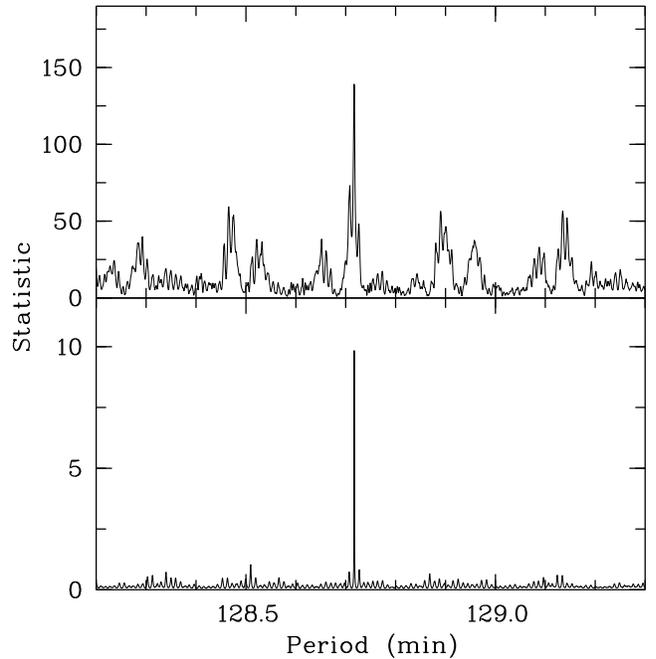}
\caption[]{\label{peri}
Periodograms  based on the analysis-of-variance method computed from all 
available optical photometry (upper panel) and least-squares method applied to 
the timings of the end of bright phase given in Table~\ref{timings} 
(lower panel). Likely periods appear as maxima.} \end{figure}

Although \117\ displayed a large variety of light curve morphologies during 
the four years of monitoring%
, it showed one persistent photometric
feature: a steep decline of at least half a magnitude. This has been 
used for a period search based on the 
analysis-of-variance method (Schwarzenberg-Cerny 1989) and a least
squares calculation applied to the barycentric timings of the 
end of the steep decline (see Table~\ref{timings}). 
The resulting periodograms (Fig. \ref{peri}) show only one
significant periodicity at 128.7~min.
 The times of the steep end of the bright hump follow a linear
 ephemeris given by 
\begin{eqnarray}
T_{\rm end}({\rm BJD}) = 2\,449\,625.5502(26) \nonumber\\
\qquad + E \times 0.089386876(10) \label{e:rx1846_eph}
\end{eqnarray}

 \begin{table}[th]
 \caption{Barycentric timings of the end of bright phase}
 \label{timings}
 \begin{minipage}{70mm}
 \begin{tabular}{rrrr}
\noalign{\smallskip} \hline \noalign{\smallskip}
\multicolumn{1}{c}{$T_{\mathrm{end}}$ } &
  $\Delta T_{\mathrm{end}}$ & O-C&\multicolumn{1}{c}{Cycle}  \\ 
\multicolumn{1}{c}{(HJD 2400000+)} & \multicolumn{2}{c}{(10$^{-4}$ d)} &  \\ 
\noalign{\smallskip} \hline \noalign{\smallskip}
49617.5070 & 25& $16 $   & $ -90$ \\
49618.4032 & 25& $39 $   & $ -80$ \\
49619.3817 & 25& $-9 $   & $ -69$ \\
49623.4024 & 20& $-26$   & $ -24$ \\
49625.4602 & 15& $-7 $   & $ -1$ \\
49625.5493 & 15& $-9 $   & $ 0$   \\
49840.8845 & 20& $ 15$   & $ 2409 $\\
49842.9370 & 20& $-18$   & $ 2432 $\\
50015.3664 & 28& $ 5 $   & $ 4361 $\\
50750.2970 & 20& $-68$   & $ 12583$ \\
50839.6037 & 14& $ 25$   & $13582 $\\
50839.6936 & 14& $ 29$   & $13583 $\\
50897.5253 &  7& $ 14$   & $14230 $\\
50897.6131 &  7& $-2 $   & $14231 $\\
51031.5133 &  7& $-13$   & $15729$ \\
51081.4827 &  7& $ 9 $   & $16288 $\\
\noalign{\smallskip} \hline \noalign{\smallskip}
 \end{tabular}\end{minipage}
 \end{table}
\noindent with the numbers in brackets giving the uncertainty in the last 
digits. The deviation of the observed times with respect to 
Eq.~\ref{e:rx1846_eph}
can be as large as 10 min, and is likely caused by an azimuthal drift of the 
primary accretion spot of the order of $30^{\circ}$.
In Fig.~\ref{f:rx1846_lc} we present the set of all available light curves
with more than one orbit coverage, folded over the ephemeris derived
above. 
Apparently,  the system dropped from a high state in 1994/1995 into an 
intermediate state in 1997/1998. 
This drop of the overall brightness of $\sim$1 mag was accompanied by a 
drastic change in the light curve morphology.

During the intermediate state (Fig.~\ref{f:rx1846_lc}, right) the light curves are 
dominated by one active pole, self-eclipsed by the body of the white
dwarf, evident by a $\sim$1 magnitude brightening every 128 min.  
The duration of the bright phase was 0.4 and 0.45 on January 25 and March 24
1998, placing the primary pole to the farther hemisphere of the white dwarf.
At these occasions the rise to maximum occurred just within a few minutes, 
and the flux remained more or less constant throughout the bright phase, showing 
no signs of cyclotron beaming. 
We also note a systematic brightening during faint phase by $\sim$0.5 mag, a 
feature which can be associated with the presence of a 
fainter secondary accretion region.
The shape of the bright phase as well as the behaviour during the 
faint phase are very similar to the well-studied self-eclipsing polar ST LMi 
(Cropper 1986), which has an inclination of $i = 56\degr$ and a
colatitude $\beta = 134\degr$.  
On August 5 1998 we observed a major restructuring of the main accretion
region. The bright phase was prolonged to 0.65 of the orbit and 
was highly asymmetric with a very slow rise to maximum for almost the entire 
time followed by a steep decline. 
This asymmetry can be understood  in terms of an accretion arc. 
The orientation of this arc has to be such, that it is orthogonal to the 
limb of the white dwarf during rise, and parallel when it disappears
behind it.

\begin{figure*}[t]
  \psfig{figure=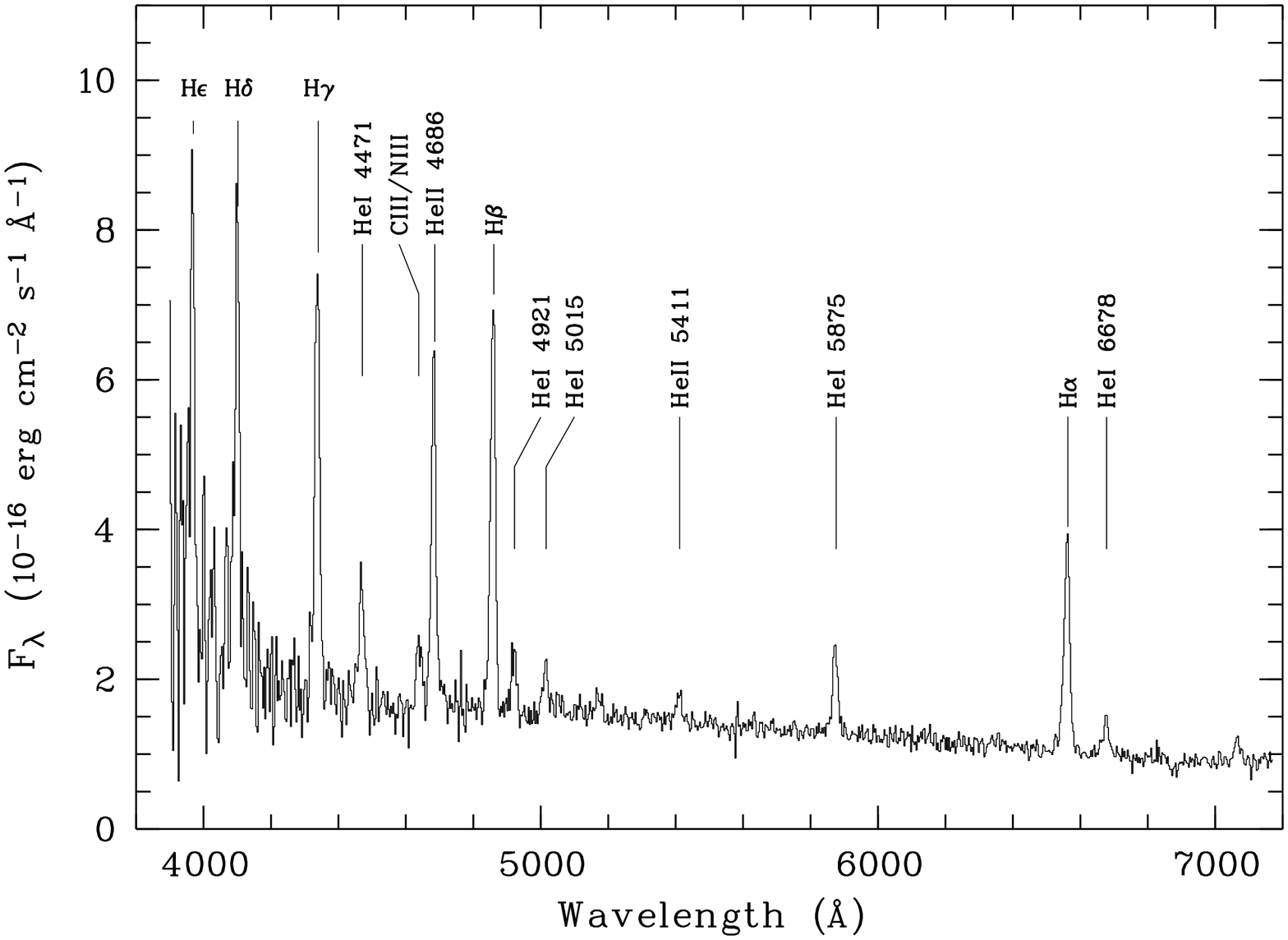,width=11.7cm,%
      bbllx=565pt,bblly=67pt,bburx=45pt,bbury=800pt,angle=-90,clip=}
  \vspace*{-8.2cm}\hspace*{11.4cm}
  \psfig{figure=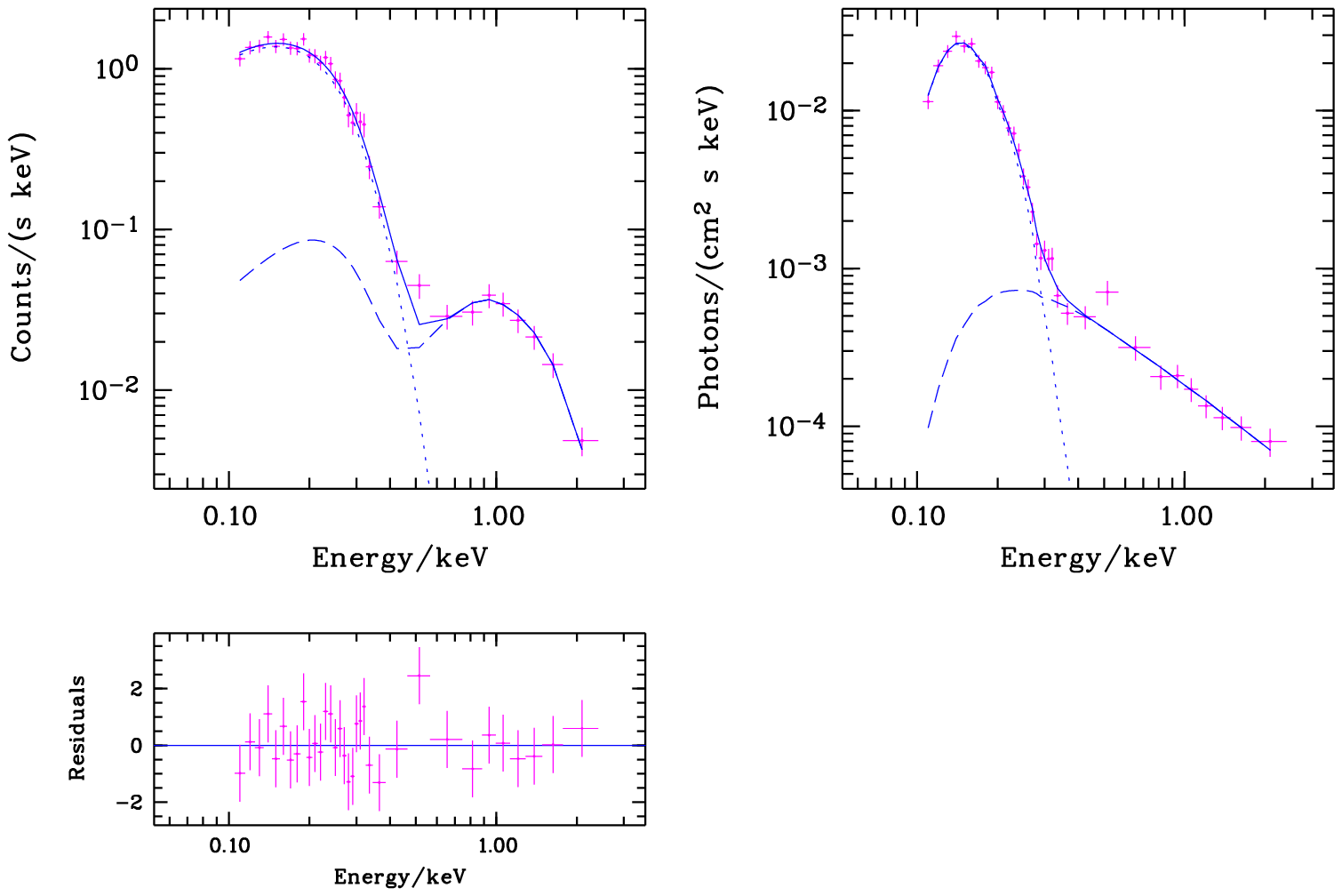,width=6cm,%
       bbllx=2.5cm,bblly=1.3cm,bburx=10.cm,bbury=11.6cm,clip=}
  \caption[osp]{{\bf Left:} Low resolution optical spectrum of \rxj\ obtained 
      on October 1, 1992. Main emission lines are indicated. 
      {\bf Right:} Phase-averaged X-ray spectrum of the June 1992 PSPC
      observation of \117\ unfolded with the sum of a blackbody and a
      thermal bremsstrahlung spectrum with a temperature fixed at 20 keV.
      The lower right panel shows the residua of the fit in units of $\sigma$.
   }
  \label{oxspec}
\end{figure*}
At the time of the optical high state the orbital variation is much less 
pronounced, but
still marked by a 0.5~mag drop at phase $\phi = 0$ due to the
disappearance of the primary accretion region. Other notable features
are  the double humped light curve on September 29, 1994, and the sharp
dips observed in May 1995 at phase $\phi = 0.7$.
The latter are coinciding in phase with the
X-ray dip seen in most of the ROSAT observations, probably caused by 
photoelectrical absorption by material in the accretion stream. 
Although less evident than the X-ray light curves, we interpret the enhanced
optical flux between $\phi = 0 .. 0.4$ as the manifestation of simultaneous 
accretion onto a second pole.

\subsection{Long-term behaviour}\label{s:rx1846_longterm}
\begin{table}
\caption{Long term variations of \117}
\begin{tabular}{cccc}
\noalign{\smallskip} \hline \noalign{\smallskip}
      Date & Type & Brightness$^{(1)}$ & Mode \\
\noalign{\smallskip} \hline \noalign{\smallskip}
    Oct 90 & X & 0.0      & one-pole \\
    Oct 91 & POSSII & $\sim20^{\rm m}$   & ? \\
    Jun 92 & X & 0.42      & two-pole \\
    Oct 92 & Spec & 18\fm 4   & one-pole \\
    Sep 93 & X & 0.05   & one-pole \\
    Sep 94 & Phot & 17\fm5   & two-pole \\
    Oct 94 & Phot & 18\fm9   & one-pole \\
    Apr 95 & X & 0.2$^{(2)}$ &  ?   \\
    May 95 & Phot & 17\fm3 &  two-pole   \\
    Oct 95 & Phot & 17\fm6 &  two-pole   \\
    Oct 97 & Phot & 19\fm6 &  one-pole   \\
    Jan 98 & Phot & 19\fm1 &  one-pole   \\
    Mar 98 & Phot & 18\fm1 &  one-pole   \\
    Aug 98 & Phot & 18\fm4 &  one-pole   \\
    Sep 98 & Phot &  &  switching   \\
    Sep 99 & Spec & 17\fm2 &  one-pole   \\
 \noalign{\smallskip}
 \hline
 \noalign{\smallskip}
      \end{tabular}\\
   \noindent{\Ni\small $^{(1)}$ Brightness between $\phi = 0..0.4$ 
                                X-ray: PSCP countrate; Optical
                                photometry: 
				magnitudes have been calculated 
				from differential white light measurements
				and the USNO A.2 R-band magnitudes of the 
				comparison star.
                       $^{(2)}$ HRI countrate has been converted using a 
                       factor of 7.8.
            }
    \label{t:rx1846_longterm}
   \end{table}
In order to inspect the frequency of the brightness and 
accretion mode changes we compiled all available measurements of
\117 (Table~\ref{t:rx1846_longterm}). 
For the single, not phase-resolved measurements 
we assigned the possible accretion mode depending on the brightness 
at phase $\phi= 0.3$, if available. 
We find in our data at least six alterations between 
single and two-pole accretion between
1990 and 1999. The amount of data  is yet insufficient to detect any possible
periodicity of these changes.  
It is noteworthy that the system was never observed in the single accretion 
state with the secondary pole being active. 

The time scale of the change is constrained  by the 
observations in September/October 1994, when \117\ dropped 
into an intermediate state within  at most two weeks.
An indication that such changes might be much faster 
comes from  the light curve 
obtained on Sep 24, 1998. At that occasion the system's brightness
at phase 0.3 faded from a level typical for the high state by 0.8 mag
within two consecutive orbital cycles.

\section{Optical spectroscopy}\label{s:rx1846_spec}

\begin{figure}[h]
      \includegraphics[width=0.98\columnwidth,clip]{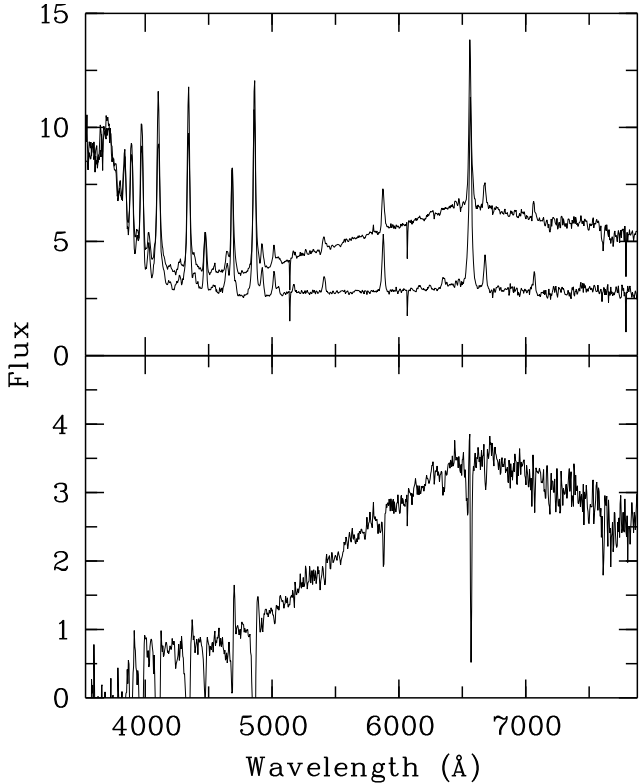}
\caption[]{\label{f:1846_saospec}
Spectroscopy obtained at SAO on June 19 1999.  In the upper panel 
averaged spectra of the bright and faint phase are displayed. 
The difference of those two spectra (lower panel) corresponds to the
cyclotron spectrum of the primary accretion region. Flux is given 
in units of $10^{-16}$\erga .
}
\end{figure}
First spectroscopic identification observations were performed on 
October 1, 1992 with the 3.5~m telescope at Calar Alto, Spain. We used
the Cassegrain spectrograph equipped with a RCA CCD as detector
covering the optical wavelength range from  3800--7100 \AA. The
observation was obtained   under stable photometric conditions and
accompanied by measurements of the standard star Feige~110, which was used
to calibrate the flux with an estimated accuracy of $\sim 20$\% (using standard
{\sc midas} procedures). By convolving the
original spectrum with functions representing the $BV\!R$ bandpasses,
we arrive at $B=18\fm6$, $V=18\fm7$ and $R=18\fm4$ mag for \117.
The identification exposure on October 1, 1992
lasted one hour corresponding to approximately half of the binary orbit, 
and was centered on HJD 244\, 8897.4271 which corresponds to 
$\phi = 0.25$ of the ephemeris given by Eq.~\ref{e:rx1846_eph}. 
Thus, the spectrum covers the complete faint phase observed in the one-pole 
accretion mode, and the magnitudes derived above are consistent 
with that found in the intermediate accretion state. 
The original spectrum is shown in Fig. \ref{oxspec}. 
It is dominated by intense
emission lines of the Balmer series, \ion{He}{II} $\lambda 4686$~\AA,
and \ion{He}{I} superimposed on a flat continuum. The inverted Balmer
decrement and the strength of the  \ion{He}{II} $\lambda 4686$~\AA\
line point to a magnetic CV classification. 

\subsection{Cyclotron spectroscopy}\label{s:rx1846_cyc}
Further phase-resolved low-resolution spectroscopy was obtained on
June 19, 1999
with the 6~m SAO telescope at Zelenchukskaja, Russia. 
The SP-124 spectrograph with a 300 l/mm grating was 
used to obtain spectra in the $3400 - 7900$ \AA\ range with 
 9 \AA\ (FWHM) resolution.  
An exposure time of 900 sec was necessary to reach a reasonable 
signal-to-noise ratio.  We obtained 14 spectra thus covering the orbital 
period twice. 
The $R$-band light curves derived from the spectra 
showed a brightening from 17$^{\rm m}$ to 16$^{\rm m}$ for half of the orbit, 
very similar to that observed in the one-pole accretion mode in 1997/98.

In order to extract the cyclotron component from the primary accretion region
we took the difference spectrum of the bright ($\phi > 0.45$) and
faint ($\phi < 0.45$) phase as shown in the lower panel of Fig.~\ref{f:1846_saospec}. 
It is characterized by a very red (zero flux below 4000 \AA) and
smooth continuum peaking at $\lambda = 6700$~\AA ,   and bears strong
resemblance with that of other low field polars like BL Hyi (15~MG;
Schwope et al. 1995a),  EP Dra (16~MG; Schwope \& Mengel 1997) or 
CP Tuc (Thomas \& Reinsch 1996). 

In this low magnetic field strength regime only the high harmonic tail ($m>
10$) is accessible in the optical, where the individual harmonics are
broadened and blended into a quasi-continuum.  The peak of the cyclotron
flux corresponds to the turnover between the optically thick part of the 
spectrum represented by the Rayleigh-Jeans tail of a black body, and 
the optically thin part, which is steeply declining towards smaller
wavelengths due to the $F\sim \lambda^{2}$ dependence of the cyclotron absorption
coefficients. The harmonic number at which the cyclotron spectrum becomes
optically thin depends on the plasma temperature and the optical depth
parameter $\Lambda$ and can be used for an estimate of the magnetic
field strength $B$ in the accretion region.

Setting the yet unknown plasma temperature to a value of 10 keV
typically found in polars we computed the expected wavelength of turnover 
for different values of $\Lambda$ and $B$ 
using cyclotron absorption coefficients 
according to 
Chanmugam \& Dulk (1981). 
For the typical range of depth parameters $log(\Lambda)=6..8$ observed in high 
accretion state polars the peak of the cyclotron flux at $\lambda =
6700$ \AA\
would correspond to a magnetic field strength in the primary accretion region 
of \117\ of $B=20..15$~MG. The integrated flux of the
cyclotron component in the range $\lambda$3400-7900 \AA\ 
is $0.6\times 10^{-12}$\ergcm . 
Accounting for a similar contribution from the infra-red not covered by
our spectra
the total, bolometric flux is of the same order as the X-ray bremsstrahlung
component.
\subsection{A rough distance estimate}\label{s:rx1846_dist}
The faint phase, intermediate state spectrum taken in October 1992 
is devoid of any spectral features of 
the secondary star (e.g. the TiO bands at $\lambda 6159$ and 6651 \AA).
We estimate that the contribution of the companion is less then 10\%  in 
the $R$-band i.e. $R_{\rm sec}\ga 19\fm9$.  This can, in principle, be used 
to derive a lower limit on the distance of the system. Assuming that the secondary in \117\
is  a dM~4.5 star, as expected for a Roche lobe filling main-sequence 
star at this orbital period, it should have an absolute magnitude of $M_{R} = 
11.89$ (Kirckpatrick \& McCarthy 1994). This leads to a distance modulus
of 8.0 or a lower limit of the distance of 400 pc. We caution however,
that the spectral features involved can be highly suppressed on the front side
of the secondary by the strong EUV/X-ray radiation. As the viewing geometry 
of \117\ is as yet unknown, we can not exclude that a non-negligible fraction of the
disturbed atmosphere visible at the time of our spectroscopy caused 
the absence of the TiO bands. This in turn would decrease the 
distance  estimated above. 

\section{Discussion}

\117 shows all major hallmarks of a cataclysmic variable of the AM Herculis
subclass: an emission line spectrum with abundant high ionisation species
like \ion{He}{II} $\lambda 4686$, and the Bowen blend, optical and X-ray light curves strongly 
modulated on the spin period due to the self eclipse of the primary accretion
region, and a very red cyclotron spectrum from that primary accretion region, 
indicating a polar field strength in the range of 15-20 MG. 
At most occasions the X-ray light curves reveal sharp X-ray dips,
likely caused by absorption within a focused accretion funnel. 
With an orbital period of $P=128.7$~min \117\ is another magnetic
CV which populates the lower edge of 
the 2--3 hr CV period gap.

\subsection{Accretion mode changes}
The rather large  body of observations available for 
\117\ reveals 
frequent switches between 
one-pole accretion and additional activity from a second,  equally
bright, pole. 
Although accretion onto a second pole has been seen in  a few 
polars, e.g. in DP Leo (Cropper et al. 1990), 
VV Pup (Wickramasinghe et al. 1989) or 
UZ For (Schwope et al. 1990), in most cases the activity 
from the secondary pole is at least one order of magnitude lower
compared to the primary one.
Remarkable exceptions include the 'anomalous' or 'reversed' state 
of  AM Her (Heise et al. 1985), 
which appeared to be a singular event, and the low accretion rate
polar HS1023+39 which permanently accretes onto two poles having similar
luminosities (Reimers et al. 1999, Schwarz et al. 2001).
So far the only polar, beside \117,
which regularly changes between the single and two-pole accretion mode
is the period-gap system QS Tel (Schwope et al. 1995a, Rosen et al. 1996, 2001).

There are two alternative mechanisms that can trigger the accretion
mode changes observed in \117. Firstly, it can be due to a slight 
asynchronism of the
spin of the white dwarf compared to the orbital motion, which
in the case of an oblique dipole results in a constantly changing orientation 
of the 
magnetic field with respect to the infalling gas stream. Matter can then
be channeled along different field lines, possibly feeding opposite magnetic
poles. While this so-called 'pole switching' is believed to operate in 
the four known asynchronous AM Herculis systems
(Campbell \& Schwope 1999), detailed studies indicate that 
the mass exchange might be more complicated. 
For example in CD Ind, the only asynchronous polar  with  
known accretion geometry, 
one primary pole dominates in the light curves 
for a large fraction of the beat cycle (Ramsay et al. 2000). 
Thus,  the fact that \117 does not switch between two magnetic poles 
but has one primary pole  might still be consistent with a possible 
asynchronism. 
A more crucial test of this scenario is the requirement that the 
accretion mode changes should occur strictly periodically over the beat
cycle, a possibility which can not yet be rejected for \117\ on the 
basis of the
available data.

The second mechanism would be variations of the total mass accretion rate
from the secondary star, which are commonly seen as low states in 
disk-less magnetic CVs. These variations are fast, and reoccur 
aperiodically on timescales of months to years.  
The working hypothesis that explains most of the observed properties
invokes the blocking of the $L_{1}$ point by starspots on the surface of
the secondary star as proposed by Livio \& Pringle (1994). 
As the total mass accretion rate increases, the ram  pressure ($\rho v$)
increases 
in the stream,  leading to a deeper penetration of the ballistic mass stream 
into the
white dwarf's magnetosphere. Possibly the stream will then connect to
field lines which can also feed the less favoured  pole. 
By now,  all observed two-pole accretion states in \117\ have been related to epochs 
of enhanced total brightness in the optical and X-ray, thus indicating 
an increase of the mass accretion rate. 
This correlation is so far the strongest argument in favour of the above
picture. 
If the pressure balance relation holds, the effective magnetospheric 
radius at which matter is controlled by the magnetic field would 
scale with the mass accretion rate with $r_{\rm mag } \propto \dot{M}^{-2/7}$. 
The brightening of the total X-ray flux by a
factor of $\sim 2$ in the two-pole state implies that the accretion rate must 
have roughly doubled at that epoch. 
The corresponding reduction of the magnetospheric radius would then be 
only of the order of 20\%, probably not enough to reach the less favoured
magnetic pole in the case of the standard field orientation seen in AM Herculis
binaries with an azimuthal angle of $\psi = 45\degr$ (see Cropper 1988).   
Indeed, long-term X-ray monitoring of the eclipsing polar HU Aqr
(Schwope et al. 2001) through different accretion states revealed only 
a moderate shift of the stagnation region of only 30\degr\  as the
accretion rate varied by a factor of 40. Thus,  in addition to the accretion
rate changes, a special field geometry is possibly required to facilitate 
the frequent accretion mode changes in \117.

Both accretion regions emit approximately the same accretion luminosities of 
$L_{\rm X}$ = 7$\times$10$^{31}$ (D / 400 pc)$^2$ erg/s, thus receiving
quite similar mass accretion rates. 
However, 
the contributions from the hot thermal plasma,  emitted as
hard X-rays,  and reprocessed radiation from the white dwarf emitted
as a blackbody in the soft X-ray band is remarkably different in both spots.
While the energy balance of the primary region
is in agreement with that of the standard shock model (Lamb \& Masters
1979), a substantial soft X-ray excess is found for the secondary spot. 
This violation has been theoretically explained in terms of 'blobby' accretion 
of high $\dot{m}$~$(> 30$ g~cm$^{-2}$~s$^{-1})$ material buried below 
the surface of the white dwarf 
(Kuijpers \& Pringle 1982). 
Observationally, a strong correlation between the magnetic field strength and 
the softening of the X-ray spectra has been found (Beuermann \& Schwope 1994) 
from a comparative study of ROSAT data. 
Probably two different processes are involved, one increasing the soft
X-ray output and the other lowering the hard X-ray component, both changing
the energy balance as observed: 
Firstly,  the specific mass accretion
rate in the accretion region will be higher for material which travels further
along converging field lines, and was consequently coupled at larger 
magnetospheric radii. Since this quantity is directly related to the surface
field strength,  for high $B$ systems the fraction of 
high $\dot{m}$ material  will be larger on average, shifting the
energy balance towards the reprocessed blackbody component (Beuermann 1998).
Secondly, for the low $\dot{m}$ part of the material, 
bremsstrahlung radiation will be suppressed due to the dominance of 
cyclotron cooling in the case of an increased magnetic field in 
the post shock plasma (Woelk \& Beuermann 1996).

Adopting this picture for \117, the X-ray softness of the secondary spot 
would be the consequence 
of  a higher field strength of $\sim 30$ MG at this pole.
Correspondingly, the energy
balance of the primary region 
would imply a lower field strength of $\sim 14$ MG, in agreement 
with the value derived from the cyclotron spectroscopy. 
As in the six polars with field measurements available for both magnetic
poles, the more active pole in \117\ would be the one with the
lower field strength. 

\subsection{Accretion geometry}

A special geometry of the mass transfer  might foster the
drastic variations of the accretion modes observed in \117. 
For example, an orientation  where the dipolar axis is perpendicular 
to the infalling stream would equally favour both poles. 
However, this is 
difficult to verify given the as yet unknown position of the
secondary star with respect to the accretion regions, and the wide range of
possible inclinations ($i\leq 70\degs$).
There is a strong indication that the primary accretion region is located
at the lower hemisphere of the white dwarf $(\beta_{1} > 90\degs)$,
from the duration of the
faint phase ($\gamma > 0.5$) measured at some occasions 
in the optical and X-ray.
Estimates for the secondary accretion region are much less constrained
due to the insufficient phase
coverage the of June 1992 observation and the overlapping visibility of
both regions around $\phi \sim
0.5$. Using the short duration of the supersoft emission ($\Delta
\phi \sim 0.3$) as a measure would place this spot on the
lower hemisphere of the white dwarf, too.

Both accretion spots are separated in azimuth  by $160\degs$, since
the two bright phases are centered at $\phi_{1} = 0.75$ and $\phi_{2} = 0.2$.
Due to the finite coupling radius the accretion regions are likely to be offset
from the magnetic pole by $\alpha \sim 10\degs$.
Therefore, the measured angular separation is consistent with both spots
accreting via one closed field line of a dipole field.
A geometry that can accomplish the all above constraints would imply 
a magnetic field axis inclined into the orbital plane. 

Another puzzling question concerns
the accretion stream dip and its 
relation to the two accretion regions.
As the azimuth of the stream dip and the primary spot are approximately the
same, one would naively expect that the absorbing matter is 
transferred to the primary accretion region.  This contradicts the 
condition for a stream occultation $i\ge \beta + \alpha$ given 
the primary spots colatitude $\beta_{1} \ge 90\degs$. 
The above conflict could be circumvented  if one allows  for
a possible vertical extent of the
ballistic part of the stream and the stagnation region,
an explanation proposed to explain a partial stream dip observed in the 
extreme UV light curves of  the high inclination polar UZ For 
(Warren et al. 1995).
It is not clear how far this is viable for the case of \117, where 
the absorbing material is denser 
and must extend much further above the orbital plane in order to produce 
a dip at the given lower inclination.

The competing scenario, absorption by material channeled to a 
second pole in the upper hemisphere, would lead to the 
occultation of X-rays from the primary region for a wide range of
possible stream geometries.
Such a view is also supported by 
the putative correlation of the density of the stream with 
the respective activity from the secondary accretion region: 
no stream dip is observed when the faint phase emission is zero (ROSAT survey),
a mildly dense dip ($N_{\rm H} \sim 10^{21}$~cm$^{-2}$) for low faint phase 
emission (Sep.~93),
and a much denser dip in the case of obvious two-pole accretion (Jul.~92). 
Further detailed studies have to show whether the footpoint of such a 
stream trajectory can be reconciled 
with the geometrical constraints for the secondary pole, 
namely the azimuthal separation to the primary spot (which would place
it on the opposite side of the white dwarf) and a high colatitude far 
from the white dwarf rotational axis.

\begin{acknowledgements}
RS would like to thank A.D.~Schwope for helpful comments on the
manuscript.
RS and JG are supported by the Deut\-sches Zentrum f\"ur
Luft- und Raumfahrt (DLR) GmbH under contract No. FKZ 50 QQ 9602\,3 and
50 OR 9106\,8. GHT acknowledges support by grants from DGAPA IN109195 and 
CONACYT No 25454-E. The ROSAT project
was supported by the German Bundes\-mini\-ste\-rium f\"ur Bildung, Wissenschaft,
For\-schung und Technologie (BMBF/DLR) and the Max-Planck-Society.

Part of Fig. 1 is based on photographic data of the National Geographic 
Society -- Palomar
Observatory Sky Survey (NGS-POSS) obtained using the Oschin Telescope on
Palomar Mountain.  The NGS-POSS was funded by a grant from the National
Geographic Society to the California Institute of Technology.  The
plates were processed into the present compressed digital form with
their permission.  The Digitized Sky Survey was produced at the Space
Telescope Science Institute under US Government grant NAG W-2166.
\end{acknowledgements}

\end{document}